# All passive architecture for high efficiency cascaded Raman conversion

**V BALASWAMY,**[1,2] **S ARUN,**[1,2] **G CHAYRAN**[1] **AND V R SUPRADEEPA**[1,*]

[1]*Centre for Nano Science and Engineering, Indian Institute of Science, Bengaluru 560012, India*
[2]*These authors contributed equally to this paper*
*\*supradeepa@iisc.ac.in*

**Abstract:** Cascaded Raman fiber lasers have offered a convenient method to obtain scalable, high-power sources at various wavelength regions inaccessible with rare-earth doped fiber lasers. A limitation previously was the reduced efficiency of these lasers. Recently, new architectures have been proposed to enhance efficiency, but this came at the cost of enhanced complexity, requiring an additional low-power, cascaded Raman laser. In this work, we overcome this with a new, all-passive architecture for high-efficiency cascaded Raman conversion. We demonstrate our architecture with a fifth-order cascaded Raman converter from 1117nm to 1480nm with output power of ~64W and efficiency of 60%.

.

## 1. Introduction

Fiber lasers based on rare-earth dopants have shown tremendous progress over the last decade. These include power scaling, reaching over 10s of kilowatts of output power, high brightness diffraction limited beams, and fully fiberized configurations etc. [1]. Such properties have made them attractive candidates in industrial, defense and basic science applications. However, the aspect of power scaling has been largely confined to Ytterbium (Yb) emission window (1050 nm to 1120 nm) due to its inherent material advantages. As an example, nearly diffraction limited beam quality Yb doped fiber lasers with an output power of 10kW in 2009 [2] and 20kW in 2013 [3] have been demonstrated. At other wavelength bands, fiber lasers are either limited in power or no suitable technology is available [1]. This is rather a serious limitation as there are number of important applications requiring high power lasers at wavelengths not covered by rare-earth doped fiber lasers. For example, high power, single mode 1.5µm fiber lasers are attractive for a variety of applications stemming from its attractive properties of eye safety and enhanced atmospheric transmission. The standard method of generating high powers at 1.5µm is by using Erbium-Ytterbium (EY) codoped fiber laser pumped at 975 nm, (where mature, low cost diode technology is available) and the highest power demonstrated was 297 W [4] for a total pump power of 1.2 kW. The slope efficiency was < 20 % at maximum power due to parasitic lasing of Yb ions. Since then, there is no further improvement in the output power using this technology. Large quantum defect (enhanced thermal load), and parasitic lasing of Yb ions at high powers are the reasons for limited power scaling. On the other hand, with Yb-free Er-doped fibers, power scaling is again limited by enhanced thermal load (large quantum defect) if 980 nm pumping is used [5] or by the reduced pump absorption and lack of efficient and cost-effective pump sources, if in-band pumping (1530nm) is used [6]. Therefore, there is a necessity to develop scalable technologies for generating high power at other wavelength regions.

Among the various nonlinear methods used for scalable generation of new frequencies, Stimulated Raman Scattering (SRS) is a promising technique as this process is naturally phase matched inside the optical fiber. And by cascading the process of SRS inside an optical fiber, high power fiber lasers can be developed at wavelengths which are otherwise inaccessible. Such lasers are called Cascaded Raman Fiber Lasers and the principle of operation of such lasers is shown in Fig 1.

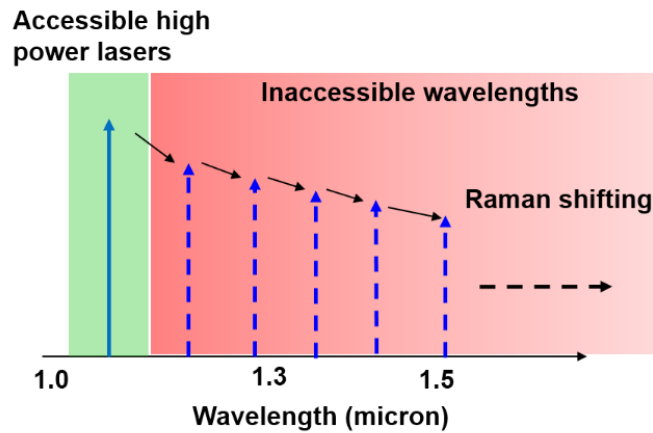

Fig. 1. Principle of cascaded Raman conversion.

In these lasers, accessible high power fiber laser sources (Yb-doped fiber lasers in this case) are wavelength converted to the required wavelength using a series of Raman shifts. Cascaded Raman Fiber Lasers provide a convenient alternative to generate scalable, high power lasers at 1.5µm and also at a variety of other wavelengths [7]. Raman lasers can also be used as convenient sources for harmonic wavelength conversion and as high brightness, low quantum defect pump sources such as those operating in the in-band absorption wavelength of Erbium doped media [8-10]. This is particularly attractive for single frequency and short pulse Erbium fiber amplifiers, where the nonlinearity is significantly reduced due to short fiber lengths made possible by high brightness pumping.

Conventional Raman lasers use a Cascaded Raman Resonator (CRR) for Raman conversion [11]. A CRR consists of a series of nested cavities at each of the intermediate stokes wavelengths comprising high-reflectivity fiber Bragg gratings and high-nonlinearity, high dispersion (to suppress other nonlinearities) fiber often referred to as Raman fiber. These lasers were limited in conversion efficiencies due to excess losses associated with the resonator assembly. Also, the stability degrades at higher powers. As the power is enhanced, coupling between CRR and rare earth doped fiber laser, results in temporal instability [12]. In order to remove the instability, an extra component, which decouples the cascaded Raman resonator from the rare earth doped fiber laser, has to be added [12]. This added extra component not only increases the losses in the system causing decrease in the efficiency but also increases the overall complexity of the system. Using this architecture, over 100W of output power was achieved at 1480nm pumped at 1117nm [11]. In this system, efficiencies were further improved through the use of Raman fiber with a long-wavelength cutoff, referred to Raman filter fiber, to suppress further conversion of final signal light to the next Stokes wavelength. Despite these additions, conversion efficiencies however, was limited to ~48 % (for a quantum limited efficiency of ~75%). Therefore, it was essential to scale the conversion efficiencies further to make cascaded Raman lasers a competitive technology.

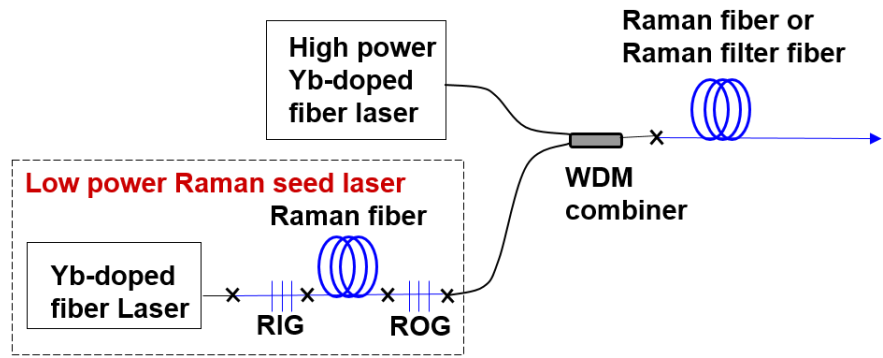

Fig. 2. Schematic of the high efficiency architecture for cascaded Raman lasers [13].

A new, high efficiency architecture, shown in Fig 2, was proposed in [13] which eliminates the need for a nested cavity. It was based on a single pass cascaded Raman amplifier, seeded at all intermediate stokes wavelengths. Seeding at all the intermediate Stokes components is essential to ensure directionality of emission, reduced gain requirements for complete conversion and narrower linewidths. Here again, the termination of the Raman cascade can be enhanced through the use of Raman filter fiber. A record output power of 301W and 1117 nm to 1480 nm conversion efficiency of 64 % was demonstrated [14]. This architecture is more reliable in terms of stability compared to conventional Raman laser architecture as temporal instability was not observed even at the highest powers.

However, in this architecture, to generate the intermediate stokes wavelengths, a complete cascaded Raman laser is needed. This necessitates an additional intermediate power fiber laser (Yb-doped fiber laser) and associated optics, increasing the overall complexity of the system. Therefore, it is strongly desired to combine both the properties (high conversion efficiency and reliability of high efficiency architecture [13, 14] and a simple, low complexity architecture of conventional Raman laser [11]), and develop a system which does Raman conversion using a single, high power fiber laser using an all-passive Raman conversion module. Such a system would be compact and cost-effective.

In our present work, we propose such a simple, all-passive architecture for high efficiency cascaded Raman conversion. We demonstrate this with a high-power, fifth-order cascaded Raman converter from 1117nm to 1480nm with output power of 64 W and conversion efficiency of ~60%. We will also demonstrate potential for generation of other wavelengths with the same system by demonstrating a 1240nm output with power of over 20W and efficiency of 62.5%.

## 2. Schematic of the experimental setup

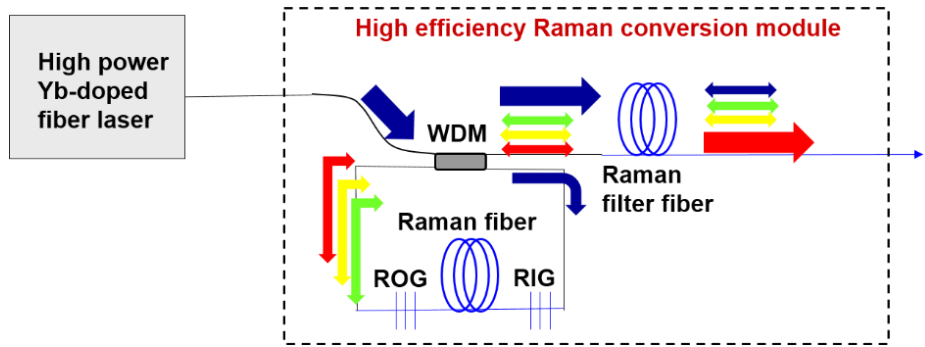

Fig. 3. Schematic of the new multi-wavelength seeded cascaded Raman amplifier, RIG, ROG – Raman input and output grating sets, WDM – Wavelength division multiplexer.

Figure 3 shows the schematic of the proposed architecture. Here, the goal of the system is to tap out the necessary fraction of power from the input high power fiber laser and use it to generate the intermediate stokes wavelengths using a conventional cascaded Raman resonator. This way, the need for an intermediate power Yb doped fiber laser required to generate intermediate stokes seed wavelengths [13, 14] is completely eliminated. The newly generated stokes wavelengths in cascaded Raman resonator also need to be efficiently coupled back with the high power laser source for cascaded Raman conversion. This necessitates a device which can tap out a small fraction of power from the high power fiber laser source while also efficiently coupling back, the newly generated stokes wavelengths. Such a functionality can be achieved with a fused fiber wavelength division multiplexer (WDM) which can operate at high power. Fused fiber WDMs with a similar intent for recombining was used in previous work in [13, 14]. In this work, their nonideality is also utilized to tap out power. If only a small fraction of power is tapped out, the hit on the efficiency is not substantial. In this work, the wavelength conversion is a $5^{th}$ order process starting from 1117nm from the Ytterbium doped fiber laser converted to 1480nm output.

In our setup, the Ytterbium doped fiber laser is a home-built system made using a linear cavity with 15m of double-clad Ytterbium doped fiber with 6/125micron core/cladding dimensions and ~1.5dB/m of peak cladding pump absorption at 976nm. The laser used a (6+1) to 1 pump combiner with 105/125 micron, 0.22NA pump ports. This laser is pumped by '3' 50W, 976nm pump diode modules providing in total ~150W of total pump power. The HR and OC gratings used in this system had bandwidth and reflectivities of (2nm, >99%) and (0.5nm, 7%) respectively. The output of the ytterbium doped fiber laser is ~107W at full power. The linewidth of the Ytterbium doped fiber laser varied linearly with output power and at full power is 2nm with a center wavelength of 1117nm. The WDM utilized in this experiment is a 1117/1480nm unit in a 2X2 configuration with an isolation between the ports of ~15dB at the design wavelengths. The transmission at all intermediate wavelengths $T(\lambda)$ followed to a high accuracy, the expression,

$$T(\lambda) = \sin^2 \theta, \theta = \frac{\pi}{2} \frac{\left(\frac{1}{\lambda} - \frac{1}{1480}\right)}{\left(\frac{1}{1117} - \frac{1}{1480}\right)}$$

The above expression is the commonly expected transmission function for fused fiber WDMs. The cross port behavior is given by $1 - T(\lambda)$. The 15-dB isolation provided by the WDM enables sufficient power to be tapped out to drive the cascaded Raman conversion. This corresponds to a tapped out power of ~3.4W at full power which is used to generate the Stokes components needed for Raman conversion in the cascaded Raman resonator. The WDM nature enables reasonable recoupling of generated intermediate stokes wavelengths. The wavelength dependent re-coupling loss of the generated Stokes components is easily overcome due to the high Raman gain in the system (of the order of 40-dB and above). In the cavity for generation of the intermediate Stokes components, 200m of Raman fiber was utilized. The Raman fiber is a conventional step-index fiber with a core-diameter of ~4micron and NA of 0.2. The small effective area results in high Raman nonlinearity and the zero-dispersion wavelength is well beyond 1800nm resulting in no instability issues related to propagation of high power light in the anomalous dispersion region. In the cascaded Raman resonator, the RIG (Raman Input grating set) and ROG (Raman output grating set) were constituted of 5 gratings at wavelengths separated by integral number of Raman Stokes shifts at 1175nm, 1240nm, 1310nm, 1390nm and 1480nm. In the case of RIG, all the gratings were high reflectivity gratings with bandwidth of 2nm and reflectivity >99%. The ROG is similar to RIG except for the 1480nm grating which is now a low reflectivity grating with ~10% reflectivity and 0.5nm bandwidth. The splice loss between the WDM and fiber laser output is ~0.5dB and the splice loss between the WDM and the Raman fiber assembly in the cascaded Raman resonator is also of the same magnitude ~0.5dB. The splice loss between the WDM and the fiber laser plays an important role in reducing the efficiency and by optimizing this, further improvement in performance can be obtained.

The Raman filter fiber used is ~120m of fiber with an effective area of 12micron^2 at 1117nm and ~16micron^2 at 1480nm. The length of the fiber was chosen based on scaled versions of previous experiments at similar input power levels [11]. The fiber is made of a W-shaped design whose linear loss is around 1dB/km until 1500nm beyond which it has a cut-off becoming non-guiding. Beyond 1520nm, the loss of the fiber increases to more than 1000dB/km. The dispersion profile of the fiber is strongly normal and < -80ps/nm/km across the bandwidth from 1117nm to 1480nm. More details on this fiber and its loss vs wavelength behavior is provided in reference [12, 13]. The output of the Raman filter fiber is angle cleaved to ~8degrees which results in suppression of any back-reflection. This ensures that no Raman converted light is seeded in the backward direction which can potentially reduce the efficiency of the laser.

We also anticipate an additional benefit from the gratings used in cascaded Raman resonator in the tap path. Though the Raman scattering is preferentially in the forward direction in this seeded architecture, it is anticipated that a fraction might be generated in the backward direction due to backward Raman scattering. In addition, distributed Rayleigh backscatter of the forward propagating components might also add to the backward component. This results in reducing the efficiency and conversion in the forward direction. However, in this architecture, as shown in the Fig 3, a substantial fraction of backward light is recoupled. In addition to reducing the fraction of light lost in the backward direction, this will also enhance the amount of seeding for forward conversion which would contribute to enhance the directionality and lower the threshold of the system.

In this work, we also observed an interesting effect. Unlike a conventional cascaded Raman resonator based laser, the cascaded Raman amplifier configuration can also be used to generate high optical powers at other intermediate Stokes wavelengths. Owing to lack of the filtering effect from the Raman filter fiber, the termination of the cascade is not as optimal in the intermediate Stokes wavelengths, however, substantial power can still be generated with good spectral quality. In this work, we will also demonstrate one such result by demonstrating a 1240nm laser with the same system described above, but run at a lower power level from the Ytterbium doped fiber laser.

## 3. Experimental results

### *3.1 Raman fiber laser at 1480 nm*

The amount of seeding generated by the cascaded Raman resonator in the tap path at full power is ~1W. Out of this, over 90% is in the final wavelength of 1480nm. We anticipate the seeding to be in the 10mW level at all the intermediate Stokes wavelengths and >0.5W at 1480nm. We anticipate the observed threshold in this system for 1480nm generation to be a function of both input power levels necessary to cascade to 1480nm and the power tapped out necessary to hit threshold in the cascaded Raman resonator. The threshold to hit 1480nm here was found to be ~1.5W. This corresponds to a minimum input power of around 47W from the input laser to generate 1480nm light for seeding.

Figure 4(a) shows the total output power and the power in 1480 nm component as a function of input power at 1117nm. We measured an output power of ~64 W at 1480 nm for an input power of ~107 W at 1117nm. The power at 1480 nm was limited by available input pump power at 1117nm and the total conversion efficiency of ~60 % was achieved (for a quantum limited efficiency from 1117 nm to 1480 nm of ~75 %). No temporal Instability was observed in the system at full power. The Raman filter fiber is single mode across the wavelength region from 1000 to 1500nm. Thus, the output of the laser is also single mode. The evolution of powers in the intermediate Stokes wavelengths as the input power is enhanced is anticipated to be similar to the previous architecture of high efficiency Cascaded Raman amplifiers. A representative figure for growth and decay of intermediate Stokes wavelengths can be found in [13].

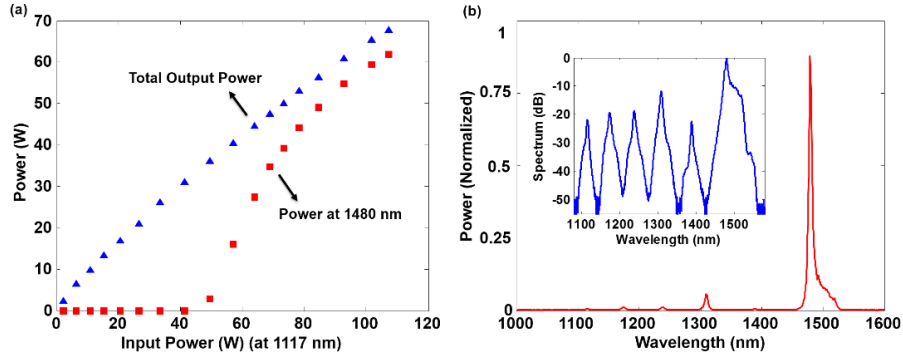

Fig. 4. (a) Plot of total output power and output power at 1480 nm as a function of input power at 1117 nm (b) Spectrum of the output in linear scale and log scale (inset) at maximum power.

Figure 4(b) shows the measured output spectrum at full power in linear scale and log scale (inset). Approximately 95 % of the power was in 1480 nm band indicating a high level of wavelength conversion and also high suppression of the next stokes order at 1590 nm is maintained through the use of filter fiber. The losses due to WDM coupler, splices, and the residual power in the other stokes components are responsible for reduction in efficiency below quantum limit. The linewidth of the final 1480nm emission is ~3nm at full power.

To illustrate the enhancement in efficiency compared to the conventional Raman laser in [11], we compare our results with those in [11] which has the same level of system complexity. The output power at 1480 nm is ~104W for a maximum input power of ~218 W at 1117nm in [11]. This corresponds to the conversion efficiency of ~47 %. But, with our current architecture, an output power of ~64 W at 1480 nm was produced with an input power of ~107 W at 1117 nm, which corresponds to the conversion efficiency of ~60%. This is a significant increase, in efficiency, over the conventional Raman lasers. However, for the same input power levels as in [11], we expect the conversion efficiencies in the proposed system to increase even more. The reason being, at higher powers, reduced length of Raman fiber is needed. This further reduces the linear loss and enhances efficiency. With the transmission loss of ~1dB/km in the fiber, there is an additional linear loss of around 3% with 120m of fiber.

To compare the performance of the current architecture with respect to the previous high efficiency architecture in [13], we compare their conversion efficiencies at similar power levels. In [13], the output power at 1480 nm is ~75 W for an input power of ~116 W at 1117 nm. This corresponds to conversion efficiency of ~62 %, which is slightly (~2%) more than the conversion efficiency in the current architecture. Thus, with this simplified architecture, we achieved conversion efficiencies similar to that in [13], but without the need of an extra intermediate power Yb-doped fiber laser.

### *3.2 Raman fiber laser at 1240 nm*

The current system is designed for operation at the 1.5 micron band. This is decided by the cut-off wavelength of the Raman filter fiber and the presence of seeding until that wavelength range. However, by suitable choice of the cut-off of the Raman filter fiber, any intermediate wavelength can be generated with this architecture. In addition, by slightly compromising on the spectral quality (fraction of power in the required final wavelength), the same system can be used to generate high powers at any intermediate Stokes wavelength. By running the input laser at a lower power level, intermediate Stokes emissions can be obtained. As a proof of concept, we demonstrate a 1240nm laser source from the same system. 1240nm sources find interesting applications as pump sources for Raman amplifiers working in the 1.3 µm band in telecommunications [15, 16].

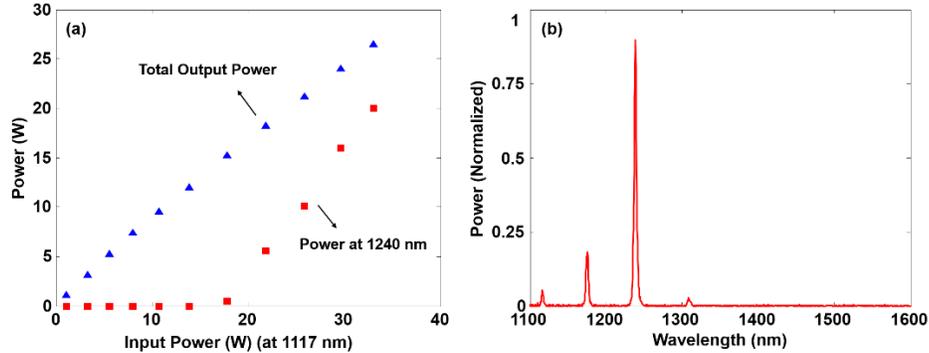

Fig. 5. (a) Plot of total output power and output power at 1240 nm as a function of input power at 1117 nm (b) Spectrum of the output in linear scale and log scale (inset) at maximum power.

Figure 5(a) shows the plot of total output power and power at 1240 nm. We measured an output power of ~20 W at 1240 nm for an input power of ~32 W at 1117 nm. The total conversion efficiency of ~62.5 % was achieved (Quantum limited efficiency is ~90%). Fig 5(b) shows the measured output spectrum in linear scale. Since the Raman filter fiber provides filtering action to at the 1500nm region, output power at 1240 nm is limited by Raman conversion to the next stokes wavelength (1310 nm). This results in sub-optimal conversion to the desired output wavelength compromising the efficiency. However, this experiment demonstrates that, with simple modifications, scalable, high power sources can be generated at a wide variety of intermediate Stokes wavelengths can also be generated.

## 4. Summary

In summary, we demonstrated an all passive architecture for high efficiency cascaded Raman conversion. We measured an output power of ~64 W at 1480 nm for an input power of ~107 W at 1117 nm, limited only by the available input power. This corresponds to the conversion efficiency of ~60% for a quantum limited efficiency of ~75% from 1117 nm to 1480 nm. The conversion efficiency is significantly higher compared to conventional cascaded Raman fiber lasers which have the same level of system complexity as the proposed system. The achieved conversion efficiency is similar to that of the high efficiency architecture proposed in [11] but with an all passive, simpler system. Thus, the proposed architecture combines the previously distinct attributes of low complexity with high efficiency. In this work, we also demonstrated that the same system can also be used to develop high optical powers at intermediate Stokes wavelengths by running the system with a lower input pump power. We demonstrated this wavelength agility through a 20W, 1240nm laser emission from the same laser system designed for the 1480 nm emission.

## 5. Funding

## 6. Acknowledgments

The authors would like to thank Jeffrey W. Nicholson from OFS laboratories for helpful discussions and OFS for providing the Raman filter fiber and Raman grating sets.